# Throughput Enhancement by Concurrent Transmission in WPAN with Multiple Antennas

Muhammad Bilal*, Moonsoo Kang*

*Department of Computer Engineering, Chosun University, Gwangju, South Korea

engr.mbilal@yahoo.com, mskang@chosun.ac.kr

*Corresponding Author: Moonsoo Kang

*Abstract*— **To achieve high rate of Multi-Giga-bits-per-second for multimedia applications at personal area level, 60 GHz communication technologies are most potential candidates. Due to some special characteristics of 60 GHz** *band of frequencies* **and use of** *multiple directional antennas,* **the network level and user level throughput can be increased tremendously by identifying and scheduling the non interfering transmission requests for concurrent transmissions. Instead of direct communication between source and destination, by traversing the traffic flow on** *optimum path* **(consists of light weight multiple-relying-hops), can further increase the throughput and balances the load condition across the network. In this paper we present a scheduling scheme for concurrent transmission of non interfering transmission requests on** *instantaneous optimum path***. Performance of scheme is investigated for different path loss exponents.**

## I. INTRODUCTION

The demand for high speed Wireless Personal Area network (WPAN) is increasing day by day. To accomplish the demand of short range high speed wireless network, in 2009 IEEE presented a standard (802.15.3c) for WPAN, which can support transmission rate greater than 2Gbps for less than 10m of range.  IEEE 802.15.3c is low cost, low power consumption, high data rate WPAN, which support the Quality of Service (QoS), easy to install and easy to manage as compare to WLANs. It defines five usage models targeting different types of high data rate demanding applications.

IEEE802.15.3c operates at 60 GHz with 9 GHz bandwidth (57~66GHz) also known as millimetre wave (mm-wave) band. IEEE802.15.3c has some especial characteristics, 60 GHz has high level of oxygen absorption, making it possible to reuse frequency in a localized region and also it uses directional antennas, which further lowering singles interference probability. These characteristics increase the probability of non interfering concurrent transmissions in localized region of WPAN. The concept of improving throughput by using concurrent transmissions has been mainly investigated for mesh networks. In case of IEEE 802.15.3c higher level of concurrency can be achieved at the cast of more complex scheduling schemes, because the identification of direction of transmission request, the identification of direction of transmission of active nodes and the accurate identification and grouping of non interfering transmission requests are crucial and difficult tasks.

By identifying non interfering nodes and scheduling them for concurrent transmissions, we can increase network and flow throughput. Because of oxygen absorption signal attenuation is significantly depends upon distance, therefore network and flow throughput also highly depends upon the distance between source and destination.

Rest of the paper organized as; in section II we briefly discuss some related work, section III describes system design consideration for network architecture, MAC layer and PHY layer, section IV explains path multihop selection mechanism, section V describes scheduling scheme and discuss PNC operation, section VI presents simulation results and finally we conclude our work in section VII.

## II. RELATED WORK

To schedule the traffic in concurrent transmissions, different scheduling schemes have been proposed so far. In [16] Maximal Weighted Matching Scheduling (MWMS) algorithm is proposed for grouping the transmission requests for concurrent transmission. The algorithm removes profound interfering edges from the traffic graph to form groups of transmission which can transmit at acceptable interfering level. The algorithm also uses weight adjustment of transmission request to change the priorities for fairness. In [4] same MWMS is investigated to check the effect of algorithm on overall energy consumption. The overall energy consumption is reduced due to making it possible to transmit in low interfering conditions. Concurrent transmission for multicast situation in multihop is discussed in [3]. The problem of scheduling is first formulated as constraint and unconstraint optimization problems then some scheduling algorithms for concurrent transmission in rate adaptive multihop networks have been proposed in [2]. The throughput analysis of single hop, single hop with concurrent transmission and multihop concurrent transmission for same topology and same source and destination has been investigated in [1]. To achieve high throughput instead of direct transmission it is better to relay the traffic on multiple hops with small distance situated between source and destination.

In this work we implemented a novel scheduling scheme for concurrent transmission and also converted the direct transmission into multihop transmissions by relaying the traffic on intermediate nodes. The path between source and destination is updated after each superframe. Finally we investigate the results for different path loss exponents and its impact on fairness.

### III. SYSTEM DESIGN CONSIDERATIONS

#### A. Network Architecture

Considering a room of X*Y square meter with set of N nodes {WT1,WT2,WT3,……WTN}, randomly deployed at different locations inside the room. All nodes are at single hop distance, means all nodes can communicate directly with each other, making a fully connected mesh topology. Each node has eight directional antennas as shown in Figure 1.

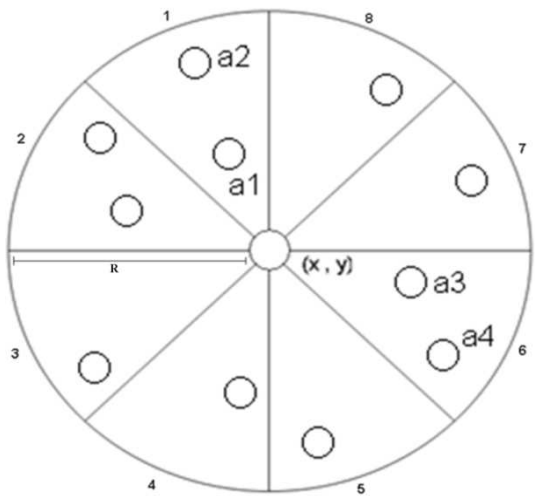

**Figure 1** Node coverage area

For example node at position (x,y) can communicate with node a1 and a2 using directional antenna "1" without interfering other nodes. But it cannot communicate with a1 and a2 at same time. Total coverage area of all antennas form a circle with radius of R meter. According to IEEE 802.15.3 any node can start a piconet and can act as a piconet coordinator (PNC).

#### B. MAC Layer

In 802.15.3c timing for data transmission is based on a superframe, which is controlled by PNC. A superframe is divided into three parts; Beacon Period (BP), Random Access Period (RAP) and Transmission Period (TP).

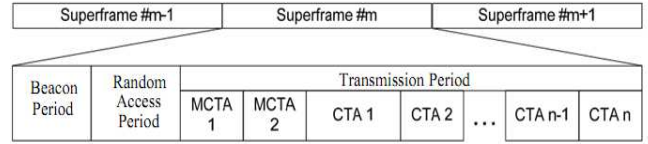

**Figure 2** IEEE 802.15.3c MAC superframe

In Beacon Period PNC broadcast synchronization and control messages. In Random Access Period WTs send requests to PNC using CSMA/CA for transmission. PNC allocates the Channel Allocation Time (CTA) or Time Slots (TS) for each request and sends the scheduling information to all nodes. During TP all scheduled nodes transmit data to next destination hop, at allocated CTA using Time Division Multiple Access (TDMA). PNC behaves as a normal node in TP and can transmit data to next destination hop, at allocated CTA.

#### C. PHY Layer

The indoor environment is less dynamic as compare to outdoor, so we can assume that channel conditions almost remain static for time duration of a superframe. In UWB throughput mainly depend upon scheduling scheme and least depend upon transmission power [2]. We can assume all nodes can transmit with maximum power.

The achievable data rate according to Shannon theory is given by;

$$R = w * \log_2[1 + SNIR] \qquad (1)$$

Where R is data rate and W is available bandwidth. In Additive white Gaussian noise (AWGN) channel SNR is given by;

$$SNIR = \frac{P_r}{(N_o + I) * W} \qquad (2)$$

Where $N_o$ is background noise, I is interference and $P_r$ is received signal power. According to Friis free space equation, path loss between two isotropic antennas is given by;

$$L = \frac{(4\pi)^2 r^2}{\lambda^2 G_t G_r} \qquad (3)$$

So, received signal power is given by;

$$P_r = \frac{P_t \lambda^2 G_t G_r}{(4\pi)^2 r^2} \qquad (4)$$

Combining equation 1,2 and 4;

$$R = W \log_2[1 + \frac{P_t G_t G_r \lambda^2}{16\pi^2 (N_o + I) W r^n}] \qquad (5)$$

Using eq (5) we calculate the channel capacity and TS requirement for a transmission request to deliver data to next hop destination. Some modification in eq (5) and calculation of time slots are discussed in section V.

## IV. PATH SELECTION

Channel capacity at 60GHz is significantly influenced by the distance, it is better to relay the traffic on multiple hops located at small distances. By considering distance as a selection metric it is possible that few nodes will suffer from heavy traffic load. To distribute the traffic across the network and to avoid any traffic bottleneck condition, [1] has introduced a metric given by following eq (6), to generate a weighted graph for all links;

$$w(i,j) = \frac{d^2(i,j)}{D^2} + \frac{f(j)}{F} \qquad (6)$$

w(i,j) is weight associated with a link between ith and jth nodes which are d(i,j) meter apart from each other and jth node (destination node in this case) has traffic load of f(j). D and F represent average distance and traffic load across the network.

Based on weighted graph, to traverse our traffic flow on optimum path, path selection between end source and destination is not made at once, rather it is dynamic. Initially path is calculated from source to destination and after TP, during RAP weighted graph is updated and next hop from requesting source node is recalculated for shortest path to destination. Which means that path between end source and destination is remain static within TP of a single superframe but it is updated and changes during RAP in each superframe. This kind of dynamic path update strategy makes it possible to select optimum path according to the most recent state of network. This also helps in distributing the workload across the network in more frequent way.

## V. SCHEDULING

### D. Transmission Requests

The scheduler receives transmission requests (TReqs) from source nodes (Source node can be any intermediate relay node) during RAP using CSMA/CA mechanism. These transmission requests are represented by a set;

TREQ = {TReq$_1$ , TReq$_2$ , ....... TReq$_k$}.

PNC represent each TReq in following form using the global information of paths and weighted graph.

- ith transmission request = TReq$_i$ < WT$_{rs}$, WT$_{nd}$, T$_{slots}$ >
- Where rs = 1,2,3.....n and nd=1,2 ,3...n. and rs ≠ nd.

Where WT$_{rs}$ is requesting source node (Source node can be any intermediate relay node), WT$_{nd}$ is next destination node on optimum path and T$_{slots}$ is the time slots requirement to send a fixed size payload of 10mb.

### E. Concurrent Transmission Group

Scheduler makes subsets of non interfering TReqs to form a concurrent transmission group (G).

G$_i$ = { TReq$_1$ , TReq$_2$ , ....... TReq$_m$ }   i=1,2,3...n
where m≤ k.

A TReq is added to the group if WT$_{rs}$ and WT$_{nd}$ of all existing TReqs are out of the range of signal interference and none of them sending or receiving data from WT$_{rs}$ and WT$_{nd}$ of new TReq. Gi and TREQ has some relational properties given below.

- G$_i$ ⊆ TREQ
- G$_i$ ∩ G$_j$ = Ø Where *i = 1,2,3.....n* and *j=1,2 ,3...n*. and *i ≠ j*.
- G$_1$∪ G$_2$∪ G$_3$∪ ... G$_i$ = TREQ Where *i = 1,2,3.....n*

### F. T$_{slots}$ Calculation

To calculate the value of T$_{slots}$ , using Shannon formula we first find out the channel capacity R by using the following modified version of eq (5);

$$R = W \log_2[1 + \frac{P_t G_t G_r \lambda^2}{16\pi^2(N_o + I * NF)Wr^n}] \qquad (7)$$

We introduce a new variable NF to adjust SNR according to active traffic flows. NF represents number of active flows within G. Due to concurrent transmission the value of I is very low but as number of active flows increases within G, the level of interference also increases.

Time slots required to send 10mb (Data payload) from WT$_{rs}$ to WT$_{nd}$ is given by,

$$T_{slots} = \frac{10/R}{t_{ts}} \qquad (8)$$

Where, t$_{ts}$ is single time slot duration.

### G. Call Admission Probability

Probability of Unsuccessful call admission has two basic reasons.
- The NLOS (within a superframe) due to moving obstacles, with a probability of 0.1[1]]
- Increase interference due to increasing number of nodes in G

We used Probability of Unsuccessful call admission given below;

$$P_{rejected} = 0.1 * size_{of(G)} \qquad (9)$$

### H. Priority Scheme

During generation of G$_i$s the highest priority is given to the TReq$_i$ with higher number of T$_{slots}$ requirement, which means giving highest priority to low rate links. This scheme is adopted to achieve fairness.

Figure-3 clearly explains the operation of PNC during a single superframe. In BP, PNC broadcasts synchronization and scheduling information. During RAP, PNC receives TReqs, PNC updates the weighted graph and calculates the optimum path. Before creating the G$_i$s, priorities are assigned to the TReqs. PNC first make a check for P$_{rejected}$, if it succeeded PNC inserts the TReq in G after checking interfering conditions and available number of time slots ( N$_{slots}$). On rejection we track the rejected TReq and we can use the track record to adjust priorities during next superframe to improve the fairness. This part is not implemented and we shown here as a future work.

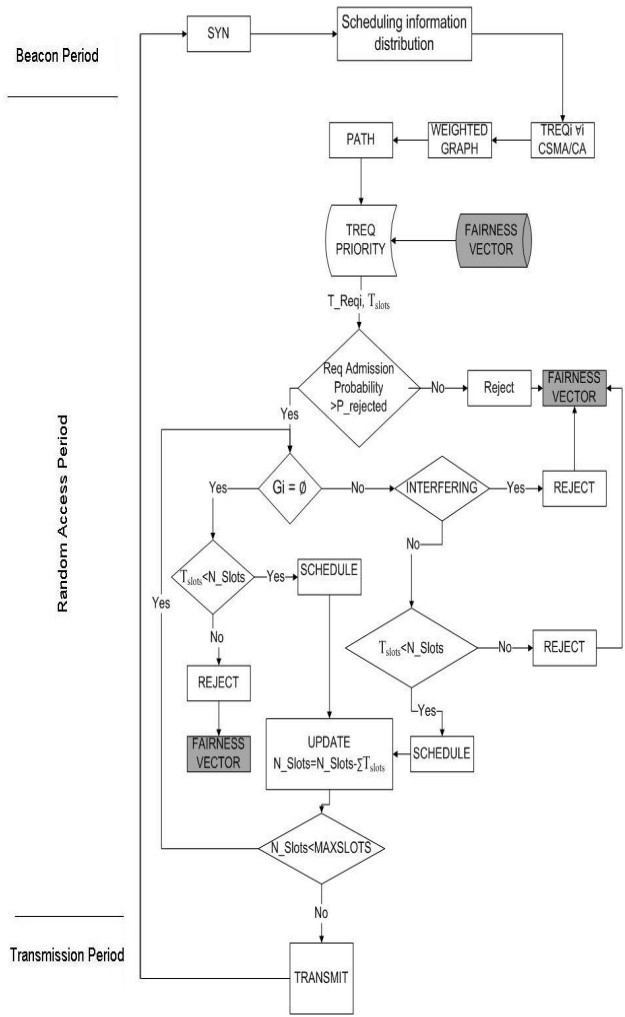

**Figure 3** Operational Flow chart

## VI. RESULTS

The simulation parameters are given in table-1.

| Parameters | Sign | Value |
|---|---|---|
| Bandwidth | W | 7000MHz |
| Transmission Power | $P_t$ | 0.1 mW |
| Antenna gain of Transmitter and Receiver | $G_t$ & $G_r$ | 12dBi |
| Background noise | $N_o$ | -134dBm/MHz |
| Path loss exponent | n | 2~3 |
| Room size | X*Y | 16x16 |
| Number of Nodes | N | 30 |

We performed extensive simulation for different number of traffic flows. The traffic pairs (Source and Destination) were selected randomly among the 30 randomly deployed nodes. For the performance evaluation we compare our Concurrent Transmission scheme (CTS) with Direct Transmission scheme (DTS). DTS doesn't traverse the traffic on small distance relay hops. In Direct Transmission when a single user scheduled for certain TS no one else can transmit during those TS, like TDMA scheme. Therefore in DTS network throughput is almost same for different number of flows.

Figure-4 shows the network throughput comparison of DTS and CTS. It is clear for different path loss exponents and number of flows, the network throughput of CTS is 100~700% higher than DTS.

Figure-5 shows the flow throughput comparison of DTS and CTS. It is clear that for different path loss exponents and number of flows, the flow throughput of CTS is 50% higher than DTS.

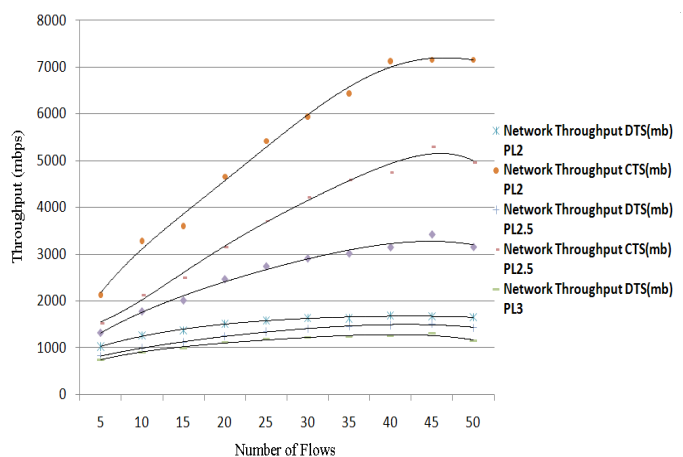

**Figure 4** Network throughput for different number of flows

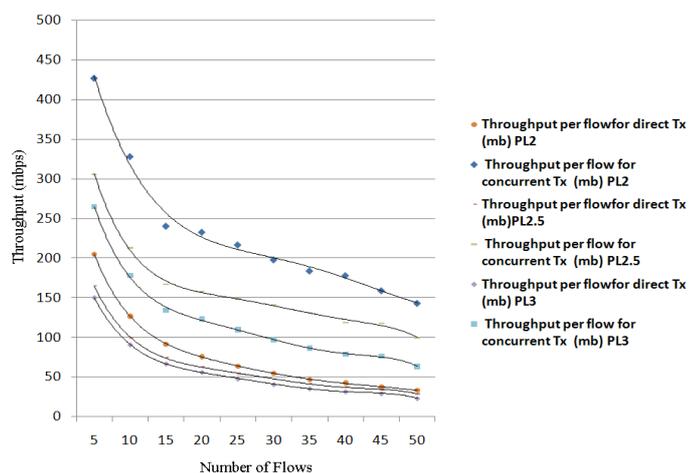

**Figure 5** Flow throughput for different number of flows

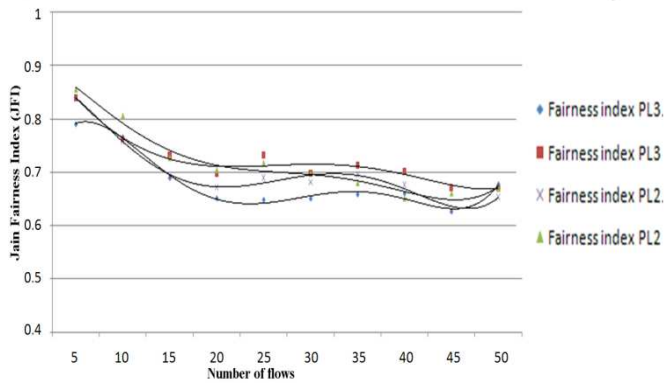
**Figure 6** Fairness index for different pathloss exponents

Figure-6 shows the Jain Fairness Index (JFI) of CTS, calculated for different Path loss exponents. Results showed that fairness is independent of Path loss exponent but it decreases by increasing number of flows. From our basic priority scheme fairness should increase with increase of Path loss exponent because we give highest priority to a TReq, with high time slot requirement, means giving highest priority to low rate link. The main reason of this behaviour is the high level concurrency is achieved with low Path loss exponent makes it possible to give fair chance to all flows for transmission.

## VII. CONCLUSIONS

Network and flow throughput can be increased by concurrent transmissions. Further improvement can be achieved by relaying the traffic on multiple hops. None line of site (NLOS) connections shows very poor performance in 802.15.3c WPAN but this problem can be solved by multihop path selection. Fairness in concurrent transmission mainly depends upon priority scheme and concurrency ratio. By increasing the concurrency fairness also increases.

## ACKNOWLEDGEMENT

This work was supported by the National Research Foundation of Korea (NRF) Grant funded by the Korea government (MEST) (no. 2011-0002405).

## REFERENCES

[1] J. Qiao, L.X. Cai and X. Shen, "Multi-hop Concurrent Transmission in Millimeter Wave WPANs with Directional Antenna," *IEEE ICC 2010 proceedings.*
[2] Z. Yang, L. Cai and W. Lu, "Practical Scheduling Algorithms for Concurrent Transmissions in Rate-adaptive Wireless Networks," i*n IEEE Proceedings INFOCOM 2010.*
[3] Z. Liu, M. Yang, H. Dai and J. Dai, "Concurrent Transmission Scheduling for Multi-hop Multicast in Wireless Mesh Networks," *IEEE International conference WiCOM, 2008.*
[4] J Shen, I. Nikolaidis and J.J. Harma," Energy–Efficient Multi–Hop Scheduling for Multi–Rate 802.15.3 WPANs," *IEEE ICC 2007 proceedings.*